\documentclass[sigconf]{acmart}
\usepackage{tikz}
\usepackage{braket}
\usepackage{pgf-umlcd}
\usepackage[edges]{forest}
\usetikzlibrary{arrows.meta,shadows.blur}
\colorlet{myyellow}{yellow!50}
\usepackage{float}
\usepackage[ruled,vlined]{algorithm2e}
\usepackage{dblfloatfix}

\usepackage[htt]{hyphenat}

\usetikzlibrary{shapes,arrows}
\usepackage{amsmath,bm,times}
\usepackage{pgfplots}
\usepackage{pgfplotstable}
\usepackage{subcaption}
\usepackage{cleveref}
\usepackage[export]{adjustbox}
\usepackage{bera}
\usepackage{listings}
\usepackage{xcolor}

\colorlet{punct}{red!60!black}
\definecolor{background}{HTML}{EEEEEE}
\definecolor{delim}{RGB}{20,105,176}
\colorlet{numb}{magenta!60!black}

\usepackage{pythonhighlight}
\definecolor{comment-text-color}{rgb}{0,0.8,0.6}
\lstdefinelanguage{json}{
    basicstyle=\normalfont\ttfamily\footnotesize,
    numbers=left,
    numberstyle=\scriptsize,
    stepnumber=1,
    numbersep=8pt,
    showstringspaces=false,
    breaklines=true,
    frame=lines,
    backgroundcolor=\color{background},
    literate=
     *{0}{{{\color{numb}0}}}{1}
      {1}{{{\color{numb}1}}}{1}
      {2}{{{\color{numb}2}}}{1}
      {3}{{{\color{numb}3}}}{1}
      {4}{{{\color{numb}4}}}{1}
      {5}{{{\color{numb}5}}}{1}
      {6}{{{\color{numb}6}}}{1}
      {7}{{{\color{numb}7}}}{1}
      {8}{{{\color{numb}8}}}{1}
      {9}{{{\color{numb}9}}}{1}
      {:}{{{\color{punct}{:}}}}{1}
      {,}{{{\color{punct}{,}}}}{1}
      {\{}{{{\color{delim}{\{}}}}{1}
      {\}}{{{\color{delim}{\}}}}}{1}
      {[}{{{\color{delim}{[}}}}{1}
      {]}{{{\color{delim}{]}}}}{1},
}

\AtBeginDocument{%
  \providecommand\BibTeX{{%
    \normalfont B\kern-0.5em{\scshape i\kern-0.25em b}\kern-0.8em\TeX}}}

\acmConference[IEEE Transactions on Computers: Special Issue on Quantum Computing]
\begin{document}

%%
%% The "title" command has an optional parameter,
%% allowing the author to define a "short title" to be used in page headers.
\title{Extending XACC for Quantum Optimal Control}

%%
%% The "author" command and its associated commands are used to define
%% the authors and their affiliations.
%% Of note is the shared affiliation of the first two authors, and the
%% "authornote" and "authornotemark" commands
%% used to denote shared contribution to the research.
\author{Thien Nguyen}
\affiliation{%
  \institution{Computer Science and Mathematics \\ Oak Ridge National Laboratory}
}

\author{Anthony Santana}
\affiliation{%
  \institution{Computer Science and Mathematics \\ Oak Ridge National Laboratory}
}

\author{Alexander McCaskey}
\affiliation{%
  \institution{Computer Science and Mathematics \\ Oak Ridge National Laboratory}
%   \streetaddress{1 Th{\o}rv{\"a}ld Circle}
%   \city{Hekla}
%   \country{Iceland}
}
% \email{mccaskeyaj@ornl.gov}

%%
%% The abstract is a short summary of the work to be presented in the
%% article.
\begin{abstract}
% Importance of quantum optimal control, we present compiler tools for optimal control in xacc, we demonstrate this capability with examples...
Quantum computing vendors are beginning to open up application programming interfaces for direct pulse-level quantum control. With this, programmers can begin to describe quantum kernels of execution via sequences of arbitrary pulse shapes. This opens new avenues of research and development with regards to smart quantum compilation routines that enable direct translation of higher-level digital assembly representations to these native pulse instructions. In this work, we present an extension to the XACC system-level quantum-classical software framework that directly enables this compilation lowering phase via user-specified quantum optimal control techniques. This extension enables the translation of digital quantum circuit representations to equivalent pulse sequences that are optimal with respect to the backend system dynamics. Our work is modular and extensible, enabling third party optimal control techniques and strategies in both C\texttt{++} and Python. We demonstrate this extension with familiar gradient-based methods like gradient ascent pulse engineering (GRAPE), gradient optimization of analytic controls (GOAT), and Krotov's method. Our work serves as a foundational component of future quantum-classical compiler designs that lower high-level programmatic representations to low-level machine instructions.
\end{abstract}

%%
%% This command processes the author and affiliation and title
%% information and builds the first part of the formatted document.
\maketitle

\section{Introduction}\label{Introduction}
Extensible and modular software architectures for quantum-classical computing are proving essential for researchers that require a quick prototyping capability and workflow customization \cite{mccaskey2020xacc, Qiskit}. Moreover, it has become increasingly evident that low-level, pulse-level access to nascent hardware will enable improved error mitigation and smart quantum-program generation technique \cite{chong_control_grape, mckay2018qiskit,2020arXiv200411205G}. There is a unique need for extensible software architectures that enable customization of the pulse-level programming, compilation, and execution workflows. 

Recently we presented a novel update to the XACC system-level quantum-classical programming, compilation, and execution framework that enables direct pulse-level programming \cite{xacc-pulse}. Specifically, we demonstrated an extension to the XACC quantum intermediate representation for pulse-level instructions, compiler scheduling routines, and an OpenPulse adherent simulation backend built on the QuaC open quantum dynamics simulator \cite{quac-github}. We believe this is foundational for the creation of higher-level quantum compiler technologies that streamline pulse-level programming research and development activities. In this work, we build off the initial XACC pulse-level software infrastructure to provide an extensible framework for typical (or custom) optimal quantum control strategies. Optimal quantum control techniques have been long-established, and a number of popular strategies exist based on typical gradient methods (GRAPE, GOAT, Krotov, etc.) \cite{PhysRevA.97.042122, PhysRevLett.120.150401, Tannor1992}. Our goal here is to present a unique software architecture that extends the XACC framework to enable the implementation of these typical control strategies in a plug-and-play manner. Ultimately, we demonstrate how this extension can be used for prototypical quantum compilation routines that lower gate-level program representations to an optimal pulse-sequence. 

We begin this work with a brief description of quantum optimal control and various gradient-based algorithms that we integrate with XACC (Sec. \ref{sec:background}). Next, we briefly provide a background on pulse-level programming and simulation in XACC (Sec. \ref{sec:xacc}), and describe in detail our extension to the IR transformation service framework enabling modular quantum optimal control strategies (Sec. \ref{Software_Arch}). We finish with a detailed demonstration of the utility of this extension for a number of problems (Sec. \ref{sec:demo}).

\section{Quantum Optimal Control}
\label{sec:background}
\begin{figure}[t!] 
\includegraphics[width=0.5\textwidth]{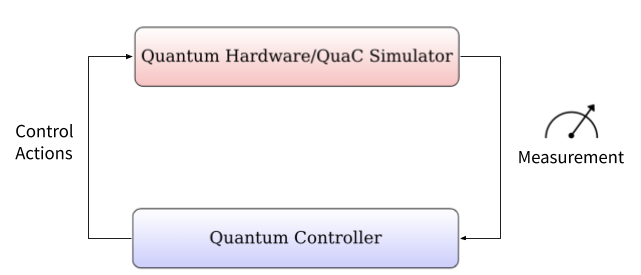}
\caption{Typical Quantum Optimal Control workflow}
\label{fig:control_diagram}
\end{figure}

To enable the control and programmability of physical qubits, one must be able to generate analog control signals that affect a desired unitary evolution with high degrees of precision. This is typically accomplished using a closed-loop feedback system with an iterative optimization algorithm acting as the controller (see Fig.~\ref{fig:control_diagram}). This task, quantum optimal control \cite{qoc-theory}, seeks to map higher-level quantum programmatic representations to an optimal sequence of pulses that realizes the unitary evolution of the program. 

% Physically, the controller outputs a waveform that is applied to the quantum system for a specified duration of time, $\tau$, usually of $\mathcal{O}(10^{-9})s.$ For single qubit systems, the control waveform is typically driven on a carrier wave that is set to the ground frequency of the qubit. Depending on the desired logic operation, the hardware, and the optimization algorithm, a user might need to fit up to three control fields to be applied over the three Cartesian axes of the system. For any given optimization algorithm, a few examples of pulse designs are:

% \begin{itemize}
% 	\item Piecewise constant functions with constraints on the amplitude \cite{article1}
% 	\item Parameterization by the mean and standard deviation of a Gaussian pulse \cite{schirmer2001quantum}
% 	\item The weighted sum of a basis function \cite{EPJQuantum.2.11, PhysRevA.97.062346} 
% \end{itemize}

% While piecewise constant functions are the easiest to solve for with small systems, they aren't robust to noise sources and timing errors, and their dimensionality scales poorly as the problem size increases. By constructing the pulse out of Gaussians, we effectively decrease the dimensionality of the optimization problem and create functions that are more suitable for physical application. Creating pulses in terms of basis functions has shown promise at designing noise-resilient control schemes \cite{PhysRevA.98.032315}, at the downside of the relative increase in parameter dimension size as compared with Gaussians. 

Quantum control algorithms typically use a loss function created out of the difference between the target state and the evolved state at time $\tau$ as a fidelity metric to guide their training. A commonly used field of algorithms are gradient-based, in that they learn the optimal controls of the system by iteratively updating them in the direction of the loss function gradient. For piecewise constant pulses \cite{article1}, this would mean iteratively updating the pulse amplitude at each time step. With Gaussian pulses \cite{schirmer2001quantum}, it would entail iterative updates to the mean and standard deviation of the distribution and with basis functions \cite{EPJQuantum.2.11, PhysRevA.97.062346}. The optimization routine is terminated once the target and evolved state reach maximum overlap with some pre-determined precision.

In this section, we seek to detail prototypical examples of gradient-based quantum optimal control strategies that we leverage in this work. 

\subsection{GRAPE}
A commonly used algorithm for quantum control is Gradient Ascent Pulse Engineering, or GRAPE \cite{PhysRevA.97.042122}. GRAPE utilizes a discretized approximation of Schrodinger's equation to create piecewise constant control pulses. The total pulse time, $\tau$, is broken up into N time-steps, all of duration $\Delta t$. GRAPE seeks to find the optimal pulse amplitude at each of the N time steps, typically subject to a value constraint on the amplitude. The time evolution of the system, in terms of its Hamiltonian, can thus be approximated as $ \mathcal{U}(t_{n}) = exp[-i \, \mathcal{H}(t_{n}) \, \Delta t] $, giving evolution at time $\tau$ as:
\begin{equation}
    \mathcal{U}(\tau) = \mathcal{U}(t_{N}) \, \mathcal{U}(t_{N-1}) \, ... \, \mathcal{U}(t_{0})
\end{equation}
The loss function that we seek to minimize, known as the infidelity, is:
\begin{equation}\label{eq:lossfunction}
    \mathcal{L} = 1 - \frac{1}{d} \Big| Tr\big[\mathcal{U}_{target}^{\dagger} \, \mathcal{U}(\tau)\big] \Big|^{2}
\end{equation}
where d is the dimension of the system Hilbert space and $\mathcal{U}_{target}$ is a user-specified target quantum gate. 

After initializing all N amplitudes, either randomly or with a pulse provided by the user, the pulse is fed to the hardware/simulator, and measurements of the system are taken. The fidelity is then recorded, and the pulse amplitudes are updated in small steps towards the direction of greatest ascent of the fidelity's gradient. The step size is a hyperparameter that may be adjusted by the user, and is used to prevent the optimizer from over or under stepping its updates. Optimization is terminated once Eq.~\ref{eq:lossfunction} is at a minima (up to some tolerance).

\subsection{GOAT}
Gradient Optimization of Analytic conTrols, or GOAT \cite{PhysRevLett.120.150401}, is another popular gradient based technique for optimal control. Unlike GRAPE, GOAT is not limited to piecewise constant controls and, therefore, can be used to create piecewise continuous pulses with bandwidth constraints. In our implementation, we optimize pulses within the GOAT framework out of a superposition of (K) Gaussian pulses of the form:
\begin{equation}
\Omega(t) = \sum_{k=0}^{K} exp(- \, (t - \tau_{k})/ \sigma_{k}^{2})
\end{equation}
where $\tau_{k}$ and $\sigma_{k}$ are the pulse duration and standard deviation of each of the K Gaussians respectively. The loss function takes the same form as in GRAPE (Eq. ~\ref{eq:lossfunction}), but with the time ordered Unitary evolution now being:

\begin{equation}\label{eq:continuous_evolution}
\mathcal{U}(\tau) = \mathcal{T} exp\Big[-\frac{i}{\hbar} \int^{\tau}_{0}\mathcal{H}(t) dt\Big]
\end{equation}

For a specified amount of pulses, GOAT seeks to optimize over the set $ \mathcal{A}_{k} = \{\tau_{k}, \sigma_{k}\}$. Beginning with either a randomly initialized or user-provided $\mathcal{A}_{k}$, GOAT uses the second-order derivative optimizer, L-BFGS \cite{MathComp35}, to iteratively minimize the loss function (Eq. ~\ref{eq:lossfunction}). The unitary evolution (Eq. ~\ref{eq:continuous_evolution}) is then computed using the third order Runge-Kutta numerical integration algorithm \cite{AppliedMath10.1016}. Training is again terminated once Eq. ~\ref{eq:lossfunction} is minimized.

\subsection{Krotov}
As opposed to GRAPE and GOAT, which use concurrent updates to optimization parameters, Krotov \cite{Tannor1992, SOMLOI199385} uses sequential updates to guarantee monotonic convergence, all without the need for gradient calculations \cite{7244489}. Application of the Krotov method requires reformulation of the system in terms of density matrices, $\rho (t)$. The system evolves in time according to:
\begin{equation}
\frac{\partial}{\partial t} \rho (t) = \frac{1}{\hbar} \mathcal{L}(t) \rho (t)
\end{equation}
where $\mathcal{L}(t)$ is the Liouvillian. For the application of quantum gates to the system, the problem may be viewed as controlling the unitary time evolution of the set of basis states, $\{ \ket{\psi_{k} (t)} \}$. Krotov seeks to minimize a functional, $J \big[ \{ \ket{\psi_{k} (t)} \}, \{ \Omega_{i} (t) \} \big]$ with constraints and boundary conditions imposed through Lagrangian multipliers, and the loss function being $\nabla_{\psi, \Omega} J = 0$ \cite{10.21468/SciPostPhys.7.6.080}. Beginning with either a randomly initialized or user-specified pulse, Krotov sequentially updates the controls by:
\begin{equation}
\Omega_{i} (t) = \Omega_{i-1} (t) + \Delta \Omega_{i} (t)
\end{equation}
until the functional is globally minimized. 

% \subsubsection{GRAFS}
% \fix{Offshoot of GRAPE, adaptation of GRAPE to constructing pulses in terms of basis function}

\section{Pulse-level Programming in XACC}
\label{sec:xacc}
XACC is a system-level software framework enabling typical heterogeneous quantum-classical computing workflows \cite{mccaskey2020xacc}. XACC can be decomposed into front-end, middle-end, and backend layers, enabling the translation of quantum kernels to a core polymorphic intermediate representation (IR), translations and optimizations on that IR, and hardware-agnostic backend execution. XACC exposes standard interfaces at all levels of this hierarchy. An interface or service of note for this work is the IR Transformation, which enables one to define general transformations on the intermediate representation. This is useful for quantum-compile time routines, and as we show in this work, mapping digital circuit representations to optimal pulse sequences. XACC has been leveraged in a number of experimental demonstrations of quantum-classical computing and general benchmarking \cite{McCaskey2019, Dumitrescu2018, PhysRevA.98.03233}. 

Recently, XACC has been updated with support for 
%With the predominance of gate model quantum programming frameworks, the XACC framework was only recently extended to support 
analog quantum programming. Here, we briefly summarize the key features of pulse-level programming in XACC, but interested readers are referred to \cite{xacc-pulse} for a more comprehensive description.

\subsection{Analog Instructions}
At its core, XACC puts forward a polymorphic Intermediate Representation (IR) as an extensible data structure that encapsulates quantum programming semantics -- from single quantum gates to composite quantum circuits. IR data structures are constructed by front-end compiler plugins, processed by middle-end IR transformation plugins (e.g., circuit optimization) and then executed on available quantum backends, e.g., remote quantum hardware or simulators.

Since the emergence of pulse-level programming, we have extended this key infrastructure of the framework to handle analog-like instructions. Specifically, we added additional fields to capture discrete pulse samples, its start time, and the target channel. Pulse instructions can then be parsed from vendor-provided pulse libraries (JSON objects), constructed manually by providing data arrays or programmatically by using a native XACC pulse generation utility which automatically discretizes commonly-used pulse shapes. 

An important aspect of pulse-level programming is the ability to automatically lower digital gates into sequences of pulses. The polymorphic hierarchy of XACC IR enables a unified representation of composite instructions, i.e. groups of other instructions (basic instructions or composite instructions), hence digital gate instructions can be replaced by a pulse composite IR which consists of multiple pulse instructions.     

\subsection{Digital-to-Analog Transpiling}
Standard IR lowering from digital to analog is included in the basic pulse extension of XACC. When used with a pulse-capable backend, such as the QuaC simulator (Sec. \ref{sec:quac}) the framework will use the backend-associated default pulse library to transpile digital gates into pulses. 

Pulse libraries typically only contain pulse sequence definitions for a pre-determined set of universal gate sets, e.g. single qubit U gates and two-qubit CNOT/CZ gates between neighboring qubits \cite{mckayZgates}. In XACC, we support a much wider range of gates. Thus, for those gates that do not have direct pulse sequence definitions, XACC transpiles them into gates drawn from the backend gate set to enable pulse conversion. Pulse sequences associated with gates are time-shifted accordingly to maintain the atomicity of quantum gates.

The result of this lowering procedure is a purely-analog composite instruction consisting of pulse instructions on different channels at different start times. This combinatorial approach toward digital-to-analog lowering is the fundamental building block of the XACC pulse programming environment upon which the quantum-control-based approach that we present here is built. Not only does quantum optimal control provide a means to derive basic pulses to construct a pulse library, but it also enables novel use cases, such as custom pulse implementation or sub-circuit optimization, which we discuss in Sec. \ref{Software_Arch}.

\subsection{QuaC Accelerator Backend}
\label{sec:quac}
In addition to a wide variety of gate-based simulation backends that are currently available in XACC, we have also implemented an OpenPulse-compatible simulation backend based on the QuaC (Quantum in C) quantum dynamics solver \cite{quac-github}. This analog backend enables users to experiment with pulse-level programming as well as to develop and verify custom digital-analog transformation procedures, e.g. those that are put forward in this manuscript.

Key components of the QuaC pulse backend are:
\begin{itemize}
\item A high-performance time-stepping solver based on the PETSc library which has built-in support for MPI parallelization \cite{petsc-web-page}. This allows us to optimize the simulator performance on platforms ranging from laptops to computer clusters.
\item An OpenPulse-compatible frontend that can process pulse-level backend information in OpenPulse format. This includes system dynamics (Hamiltonian and qubit dimensionality), drive/control channel configurations, and pulse library.
\item A pulse generation utility which supports automatic discretization of analytical pulse envelopes.
\end{itemize}

The QuaC pulse backend implements the standardized \texttt{Accelerator} interface of XACC, hence it can be used as a drop-in replacement for any other existing gate-based backends as well as in the Pythonic programming environment.

\section{Software Architecture} \label{Software_Arch}
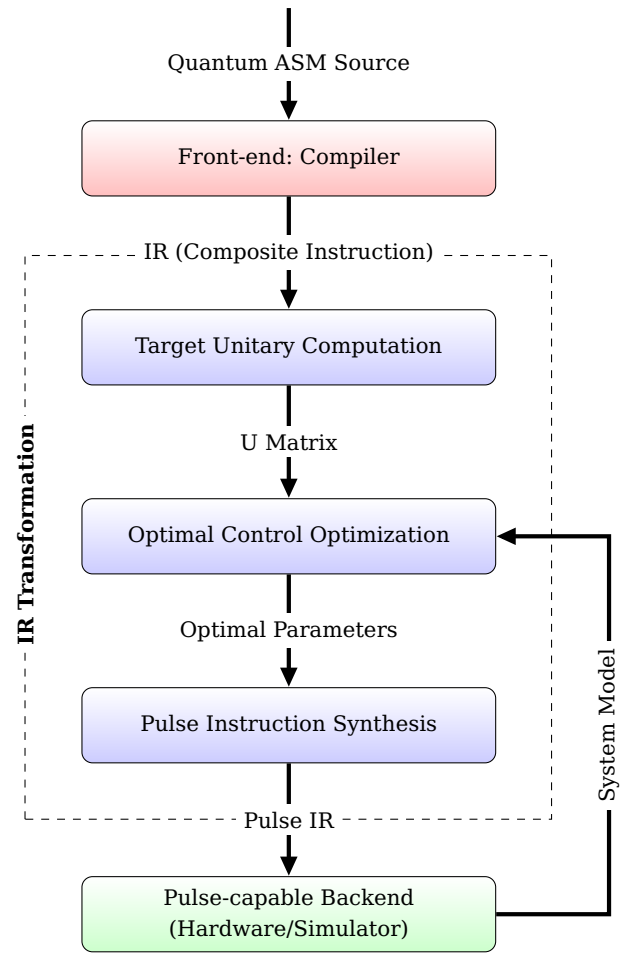
\begin{figure}[b!]
\usetikzlibrary{shadows.blur,positioning,calc,arrows.meta}
\tikzstyle{pinkbox}=[draw,rounded corners,shade,top color=white,bottom color=pink,
                     minimum width=5.5cm,minimum height=1cm,align=center,node distance=1.5cm]
       
\tikzstyle{bluebox}=[draw,rounded corners,shade,top color=white,bottom color=blue!20,
          minimum width=5.5cm,minimum height=1cm,align=center,node distance=1.5cm]

\tikzstyle{greenbox}=[draw,rounded corners,shade,top color=white,bottom color=green!20,
          minimum width=5.5cm,minimum height=1cm,align=center,node distance=1.5cm]

\tikzstyle{inter arrow}=[->,ultra thick,-{Triangle[angle=45:8pt]}]   
\begin{tikzpicture}
\node (A)[pinkbox] {Front-end: Compiler};
\node (B)[bluebox,below=of A] {Target Unitary Computation};
\node (C)[bluebox,below=of B] {Optimal Control Optimization};
\node (D)[bluebox,below=of C] {Pulse Instruction Synthesis};
\node (E)[greenbox,below=of D] {Pulse-capable Backend \\ (Hardware/Simulator)};

\draw [dashed] (D.west) + (-0.75, -1.25) rectangle ++(6.25, 6.25);
\node[fill=white] at (-3.5,-5) {\rotatebox{90}{\textbf{IR Transformation}}};

\draw[inter arrow] ($(A)+(0,2)$)--node[fill=white]{Quantum ASM Source}(A);
\draw[inter arrow] (A)--node[fill=white]{IR (Composite Instruction)}(B);
\draw[inter arrow] (B)--node[fill=white]{U Matrix}(C);
\draw[inter arrow] (C)--node[fill=white]{Optimal Parameters}(D);
\draw[inter arrow] (D)--node[fill=white]{Pulse IR}(E);
\draw[inter arrow] (E.east)--+(1.5,0)|-node[fill=white,pos=0.25]{\rotatebox{90}{System Model}}(C);
\end{tikzpicture}
\caption{XACC pulse level IR transformation flow.}
\label{fig:pulse_ir_transformation_flow}
\end{figure}

In the XACC framework, quantum optimal control capabilities are tightly integrated into the end-to-end programming model, as illustrated in Fig.~\ref{fig:pulse_ir_transformation_flow}. More specifically, with the core intermediate representation (\texttt{IR}) providing a universal data structure for describing both digital and analog quantum instructions, we are able to encapsulate quantum optimal control strategies as general transormations of the IR, specifically via a new pulse-level IR Transformation service providing a flexible gate-to-pulse lowering/compilation procedure.
\begin{figure*}[t!] 
\resizebox{\textwidth}{!}{
\begin{tikzpicture}
    \begin{interface}{IR Transformation}{0,0}
        \operation[0]{+ name()} 
        \operation[0]{+ type()}
        \operation[0]{+ apply(CompositeInstruction, Accelerator, Options)}
    \end{interface}
    
    \begin{class}{Circuit Optimizer}{-3,-4} \implement{IR Transformation}
        \operation[0]{+ name() = `circuit-optimizer'}
        \operation[0]{+ type() = `Optimization'}
        \operation[0]{+ apply(...)}
    \end{class}

    \begin{class}{Circuit Placement}{-3,-8} \implement{IR Transformation}
        \operation[0]{+ name() = `default-placement', `rotation-folding', `swap-shortest-path'}
        \operation[0]{+ type() = `Placement'}
        \operation[0]{+ apply(...)}
    \end{class}

    \begin{class}{Pulse Transform}{3,-4} \implement{IR Transformation} 
        \operation[0]{+ name() = `quantum-control'}
        \operation[0]{+ type() = `Optimization'}
        \operation[0]{+ apply(...)}
    \end{class}
    
    \begin{class}{QuaC Accelerator}{8,0.0} 
        \attribute{System model}
        \operation{execute(CompositeInstruction)}
    \end{class}

    \draw[umlcd style dashed line,->] (Pulse Transform) --node[above, sloped, black]{$<<$reference$>>$} (QuaC Accelerator);

    \begin{interface}{Optimizer}{15,0}
        \operation[0]{+ name()} 
        \operation[0]{+ setOptions()}
        \operation[0]{+ optimize(OptFunction?)}
    \end{interface}
    \begin{class}{ControlOptimizer}{11,-4} \implement{Optimizer}
        \operation[0]{+ name() = `quantum-control'} 
        \operation[0]{+ setOptions(\{`method' : name, ...\})}
    \end{class}
    \begin{class}{PulseOptGOAT}{20,-5} \implement{Optimizer}
        \operation[0]{+ name() = `GOAT'} 
        \operation[0]{+ setOptions(\{`target-U' : matrix, ...\})}
    \end{class}
    \begin{class}{PulseOptGRAPE}{19,-8} \implement{Optimizer}
        \operation[0]{+ name() = `GRAPE'} 
        \operation[0]{+ setOptions(\{`target-U' : matrix, ...\})}
    \end{class}
    
    \draw[umlcd style dashed line,->] (Pulse Transform) --node[above, sloped, black]{$<<$instantiate$>>$} (ControlOptimizer);
    
    \draw[umlcd style dashed line,->] (ControlOptimizer) --node[above, sloped, black]{$<<$instantiate$>>$} (PulseOptGOAT);

    \draw[umlcd style dashed line,->] (ControlOptimizer) --node[above, sloped, black]{$<<$instantiate$>>$} (PulseOptGRAPE);

    \begin{object}[text width=6cm]{PyOptimizer}{11, -9}
        \instanceOf{Optimizer}
        \operation{def \_\_init(self)}
        \operation{def name(self)}
        \operation{def setOptions(self, options)}
        \operation{def optimize(self)}
    \end{object}
    \umlnote (note) [text width=6cm] at (11, -12) {Python pulse optimization packages};
    \draw[umlcd style dashed line,->] (ControlOptimizer) --node[above, sloped, black]{$<<$instantiate$>>$} (PyOptimizer);

\end{tikzpicture}
}
\caption{XACC pulse-level IR transformation software architecture: single service entrance via the Pulse Transform plugin; the specific optimal control method to be used for circuit-to-pulse transformation is provided as an option; both native (C++) and Python pulse optimization methods can be invoked via this unified API.}
\label{fig:sw_arch_diagram}
\end{figure*}
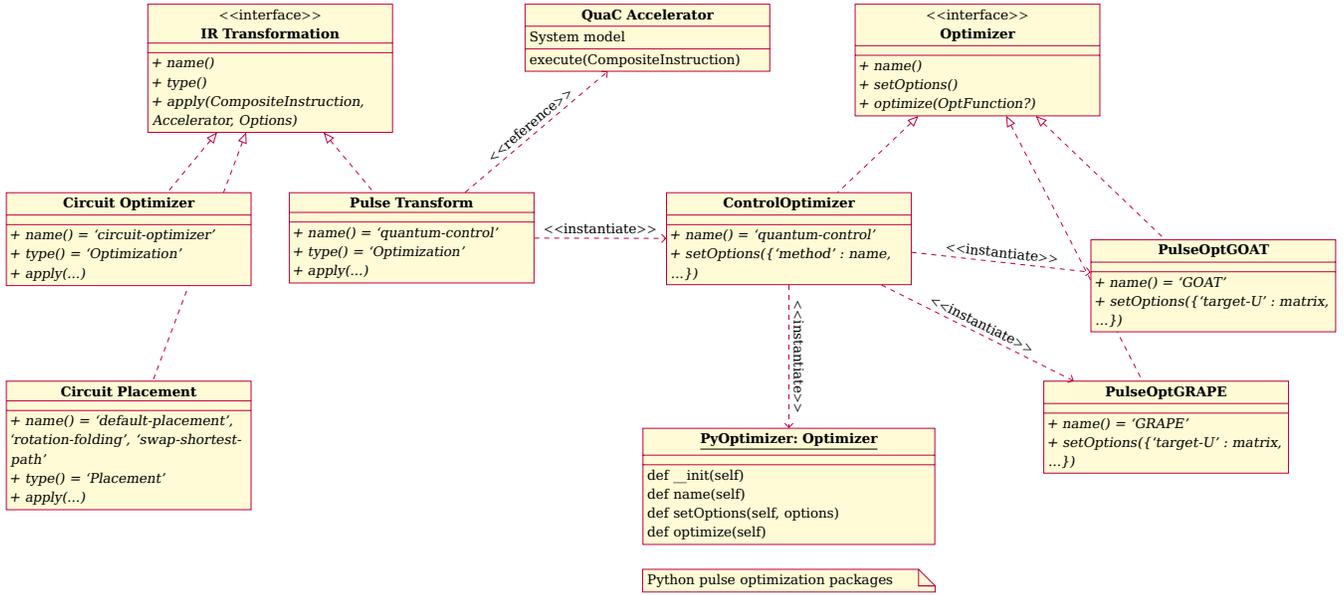
Since the input IR can represent individual gates as well as whole quantum circuits, e.g. variational ansatz state-preparation, Quantum Fourier transform, etc., this new pulse-level transformation infrastructure of XACC can provide substantial improvements in areas such as execution fidelities and efficiencies (total time of gate operation).

It is worth noting that, while quantum optimal control modules are utilized internally within the pulse-level IR transformation workflow, they can also be used independently as a service. For example, one can manually define the target unitary and the system definition and then invoke any quantum optimal control module that the framework provides.

In the following, we will describe the three main components of the pulse-level IR transformation workflow, as illustrated in Fig.~\ref{fig:pulse_ir_transformation_flow}, in greater detail.  

\subsection{IR Transformation}\label{sec:pulse-ir-transform}
Built upon the concept of compiler optimization routines from classical computing, XACC defines the \texttt{IRTransformation} interface as the backbone of the middle-end pipeline. This allows modular and customizable multi-pass transformations of quantum kernels parsed by the front-end. As shown in Fig.~\ref{fig:sw_arch_diagram}, we already have built-in support for several gate-based transformations such as gate optimization and qubit placement. 

In this work, we provide a new \texttt{IRTransformation} service called \texttt{PulseTransform}, which bridges digital and pulse \texttt{IR}'s via quantum optimal control. When invoked with an \texttt{apply()} call, the \texttt{PulseTransform} service is provided with a \texttt{CompositeInstruction} describing a quantum circuit and an instance of a pulse-capable back-end, such as the XACC QuaC simulator (see Fig.~\ref{fig:pulse_ir_transformation_flow} \& \ref{fig:sw_arch_diagram}). The backend supplies the system dynamics information which is required by downstream control modules to compute analog driving signals.

One key functionality that the top-level \texttt{PulseTransform} service performs is to convert arbitrary gate-based circuits into their equivalent unitary matrix. Thus, the framework can transform either individual gates or multi-gate circuits into monolithic pulse programs representing the underlying total unitary evolution. 

As described in the next section, we also implement a wide variety of quantum optimal control modules as well as provide a user-friendly interface to integrate custom pulse optimizers, any of which can be used in this digital-to-analog IR transformation pipeline. By specifying the method name in the input \texttt{HeterogeneousMap} of the \texttt{apply()} call, the corresponding optimal control plugin will be delegated (Fig.~\ref{fig:sw_arch_diagram}) to perform the optimization task. Also, the computed target unitary matrix along with system definitions are sent on to downstream optimization plugins as needed.

\subsection{Quantum Control Optimizer}\label{sec:quantum-control-optimizer}
The QCOR language specification \cite{mintz2019qcor} put forward the \texttt{Optimizer} data structure which provides a common interface for all classical optimization services, and XACC has provided the first definition and implementation of it. In a conventional optimization setting, an \texttt{Optimizer} implementation will perform a multi-dimensional optimization of a target cost function ($argmin_{x \in \mathbb{R}^n}f(x)$). This function is listed as the (input \texttt{OptFunction} parameter of the optimize method of the \texttt{Optimizer} interface in Fig.~\ref{fig:sw_arch_diagram}. 

The \texttt{Optimizer} interface is thus amenable to quantum optimal control problems underpinning analog control synthesis. For instance, one can consider control values at each discrete time step as parameters to be optimized and use any fidelity measures, such as the trace distance, as the function to be optimized. To maximize flexibility, modularity, and reusability of these quantum optimal control sub-routines, we design a two-level interface for the quantum optimal control service.

At the top-level, we define a \texttt{ControlOptimizer} service which can be invoked by its generic name, \texttt{quantum-control}. The caller then provides a \texttt{HeterogeneousMap} which contains a method field indicating which optimal control method to be used along with an arbitrary set of additional parameters. The \texttt{ControlOptimizer} will then look up in the XACC service registry for the specified optimal control module, initialize it with method-specific parameters (either user-provided or default values customized for each method), and delegate the \texttt{optimize()} call to the concrete implementation provided by each optimal control module.

At the time of writing this manuscript, we have implemented the \texttt{GOAT} and \texttt{GRAPE} pulse optimization methods within the XACC framework as native plugins. Hence, these two methods are available universally on any XACC installations. We also provide a Python binding interface (see \texttt{PyOptimizer} in Fig.~\ref{fig:sw_arch_diagram}) through which one can wrap Pythonic quantum optimal control modules, e.g. QuTip \cite{qutip}, and contribute them as services to be used in the XACC IR transformation workflow. In the demonstration section, we will demonstrate the use of the Krotov package, which depends on QuTip, as a backend optimizer for a pulse-level IR transformation.   

\subsection{Pulse Instruction Synthesis}\label{sec:pulse-synthesis}
As illustrated in Fig.~\ref{fig:pulse_ir_transformation_flow}, the final output of the pulse-level \texttt{IRTransformation} is a pulse \texttt{CompositeInstruction} which contains pulse instructions. Those analog instructions are defined in terms of arrays of complex-valued samples representing control signals at a back-end specific sampling rate of $dt$. Depending on the specific optimal control method that was used in the previous step, the results may not immediately be in the right format. 

For example, analytical optimization approaches, such as GOAT, produce a set of functional parameters representing time-continuous signals. In that case, the Pulse Transform service will evaluate those envelope functions and generate the corresponding data samples to construct output pulse instructions. On the other hand, time-series based methods, e.g. piecewise constant optimization, should already generate the optimized pulses as sample arrays which can be used to construct pulse \texttt{IR}'s directly.

\section{Demonstration}
\label{sec:demo}
% \fix{simple goat and grape, one qubit, two qubit, perhaps 4 qubit chemistry?, ML DRL example, qutrits, you guys add any other ideas here
%Can show a figure of gate fidelity vs optimization iteration for each of the algorithms here. Maybe do one of GRAPE vs. GOAT vs. DRL on a single qubit, then do one of DRAG vs DRL on a qutrit. 
%Alternatively, could focus this as a demonstration of our software vs other platforms like Qutip and Q-Control. Show how easy it is to set up the control problem in XACC and then plug and play with all of the optimizers that we %offer. }

\subsection{ControlOptimizer API}

In this example, we show how the underlying quantum control optimizer plugins can be used directly. This type of usage fits physics-based experiments whereby the system dynamics, target unitary, and method-specific parameters need to be provided and are fully customizable by users. 
\begin{figure}[t!] 
\begin{python}
# Run Quantum Optimal Control GOAT method to 
# find gaussian pulse that approximates the X
# gate on qubit 0
goat = xacc.getOptimizer('quantum-control', {
    'method':'GOAT', 
    'dimension': 1, 
    'target-U':'X0',
    'control-params':['sigma'],
    'control-funcs':['exp(-t^2/(2*sigma^2))'],
    'control-H':['X0'],
    'max-time': 100,
    'initial-parameters':[8.0]
})

optimal_sigma = goat.optimize()[1][0]
\end{python}
\caption{Using the \texttt{ControlOptimizer} service, i.e. the XACC Optimizer plugin named ``quantum-control", to optimize for a $\pi$-pulse (X gate) using the GOAT method. Since this is an analytical method, we need to provide the functional form of controls and the \texttt{optimize()} function will return the optimal parameters for those input control functions.}
\label{fig:simple_goat_snippet}
\end{figure}

Fig.~\ref{fig:simple_goat_snippet} is a Python snippet demonstrating the way that underlying quantum optimal control modules (plugins) are invoked. As shown in Fig.~\ref{fig:sw_arch_diagram}, all of those modules are sub-components of the high-level \texttt{ControlOptimizer} which is a generic XACC Optimizer with name ``quantum-control" (Fig.~\ref{fig:simple_goat_snippet}, line 4.)

In this example, we request the \emph{GOAT} (\textbf{G}radient \textbf{O}ptimiz-\\ation of \textbf{A}nalytic con\textbf{T}rols) optimizer by specifying its key in the \texttt{method} field (Fig.~\ref{fig:simple_goat_snippet}, line 5.) The XACC \texttt{HeterogeneousMap} utility allows us to pass flexible data-structures in a type-safe manner to the underlying native (C++) module. This makes the Python-C++ integration seamless as demonstrated in this example. 

Those control parameters after the \texttt{method} field in the \texttt{getOptimizer()} call are method-specific. For example, since GOAT is an analytical method, users need to specify the functional form of control envelopes. It is also worth pointing out that due to the fact that this is a direct invocation of an optimal control module, there are quite a few parameters that need to be specified, such as those related to the system dynamics (Hamiltonian) and the target unitary. When these underlying optimal control modules are invoked within the \texttt{IRTransformation} workflow (Fig.~\ref{fig:pulse_ir_transformation_flow}), most of these parameters will be derived automatically by the high-level \texttt{IRTransformation} service.

\subsection{PulseTransform API}

As a system-level framework, optimal control modules within XACC are tightly integrated with the end-to-end compilation, transformation, execution software workflow. Fig.~\ref{fig:simple_ir_transform_grape} demonstrates the use of our built-in \emph{GRAPE} optimizer to derive an optimal pulse shape that implements a quantum circuit. The input circuit is given as an assembly source (XASM dialect) which is compiled into digital \texttt{IR} by XACC's \texttt{Compiler} service.
\begin{figure}[t!] 
\lstset {language=C++}
\begin{lstlisting}
// Get QuaC accelerator (pulse-capable backend)
// Note: systemModel contains Hamiltonian and channel configs.
auto quaC = xacc::getAccelerator("QuaC", { 
  std::make_pair("system-model", systemModel), 
});    

auto xasmCompiler = xacc::getCompiler("xasm");
// Using the XASM compiler to compile ASM to IR
auto ir = xasmCompiler->compile(R"(__qpu__ void circ(qbit q) {
  H(q[0]);
})", quaC);

auto program = ir->getComposite("circ");

// Pulse IR transformation configs, using GRAPE:
xacc::HeterogeneousMap configs {
  std::make_pair("method", "GRAPE"),
  std::make_pair("max-time", 10)
};
// Get the pulse-level IR Transformation service
auto opt = xacc::getIRTransformation("quantum-control");
// Apply the transformation on the gate-level program 
opt->apply(program, quaC, configs);
// After the transformation, the program is converted 
// optimal pulse instructions which 
// can be simulated on the QuaC backend
auto qubitReg = xacc::qalloc(1);    
quaC->execute(qubitReg, program);
\end{lstlisting}
\caption{Using the pulse-level \texttt{IRTransformation} service to convert a quantum circuit, in this case, just a Hadamard gate, into an optimal pulse instruction using the GRAPE optimizer.}
\label{fig:simple_ir_transform_grape}
\end{figure}

This digital gate-based circuit, in this case, just a single quantum gate, is given to the pulse-level \texttt{IRTransformation} service along with a reference to a pulse-capable backend and a set of configuration parameters (see line 23, Fig.~\ref{fig:simple_ir_transform_grape}.) In this example, we want to request the gate-to-pulse transformation via the GRAPE method, which only requires a single parameter (maximum time horizon, \texttt{max-time}.) The rest of the necessary parameters, e.g. the number of data samples (GRAPE is a time-series based method), the Hamiltonian, the target unitary, etc. , are deduced by the \texttt{IRTransformation} service which has access to the execution backend and the \texttt{IR} of the digital quantum circuit.

The input digital program is lowered to a corresponding analog program at the end of the \texttt{apply()} procedure. This pulse program can then be executed on the backend as shown in lines 27-28 of Fig.~\ref{fig:simple_ir_transform_grape}.

\begin{figure}[b!] 
\includegraphics[width=0.5\textwidth]{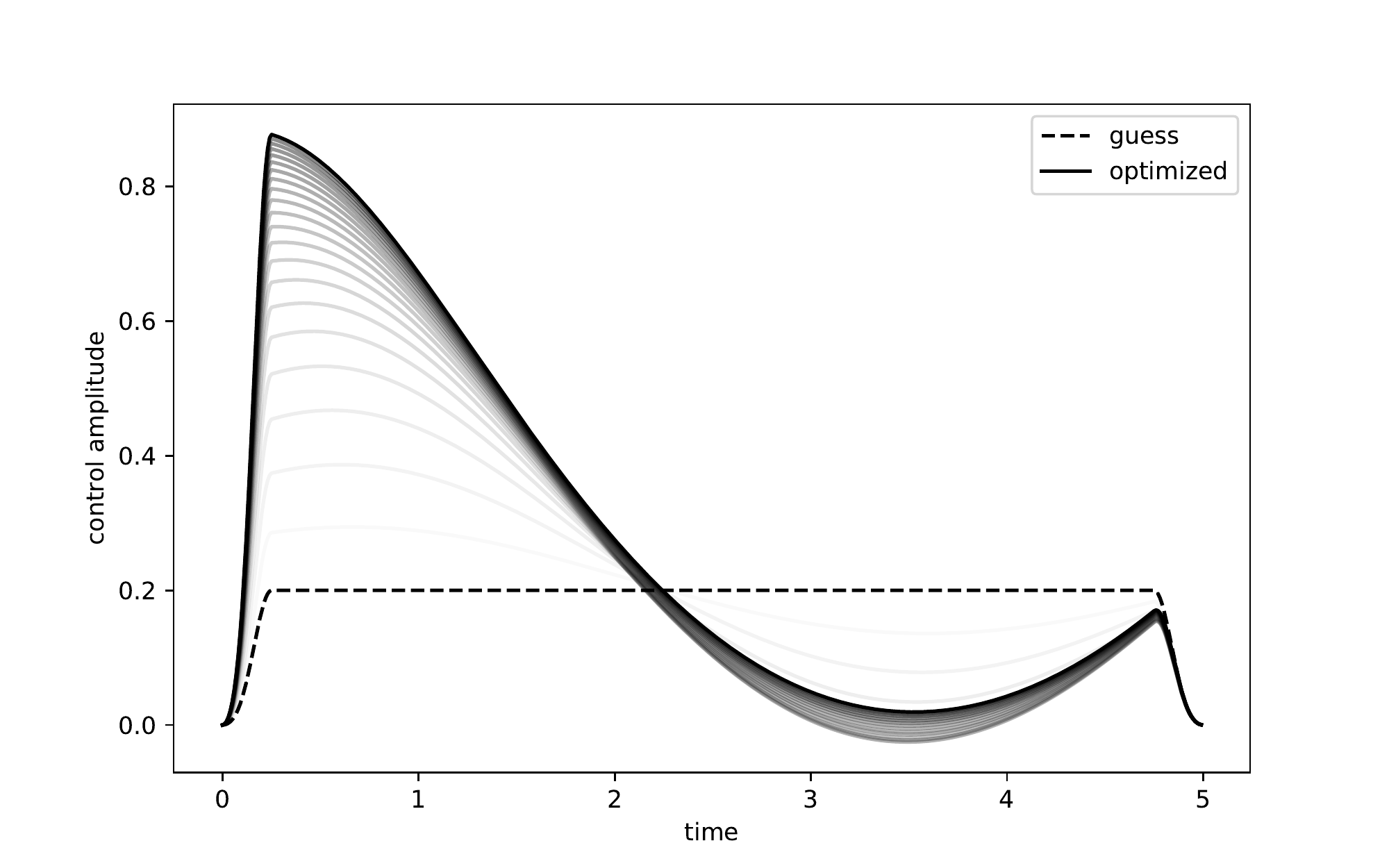}
\caption{Optimal control signal generated by the Krotov pulse optimizer plugin when invoked by the XACC \texttt{IRTransformation} service to optimize for an Hadamard unitary. The dash line is the initial pulse (randomly selected) and the blurred lines represent optimization iterations until it converges to the optimized pulse (the solid line.)}
\label{fig:krotov_control_field}
\end{figure} 
\subsection{External Python Package Integration}
In an effort to take advantage of quantum optimal control modules which are currently available either as open-source software, e.g., QuTiP, or as commercial solutions, e.g. Q-CTRL \cite{qctrl-paper}, the XACC framework provides a set of Python bindings which can be used to wrap external Python modules as plugins (see Fig.~\ref{fig:sw_arch_diagram}.) 
The primary task of the wrapper is to translate standardized data which the \texttt{IRTransformation} plugin sends on to the user-requested quantum control method into the implementation-specific data format. For instance, after reading the input digital circuit, the \texttt{IRTransformation} module will generate the target unitary matrix represented by a complex-valued vector. This bare array may need to be marshaled into the expected data format of the external Python package.

As an example, we have implemented a wrapper for the \emph{Krotov} package which depends upon QuTiP. From the top-level, i.e. \texttt{IR Transformation}, the usage is completely analogous to other built-in optimal control modules as can be seen in Fig.~\ref{fig:krotov_snippet}.  
The specific method key, in this case, `\texttt{krotov}' (line 14), is registered by the wrapper service.  
\begin{figure}[t!] 
\begin{python}
# Get the XASM compiler
xasmCompiler = xacc.getCompiler('xasm');
# Composite to be transform to pulse: H gate = Y^(1/2) - X
ir = xasmCompiler.compile('''__qpu__ void f(qbit q) {
    Ry(q[0], pi/2);
    X(q[0]); 
}''', qpu);
program = ir.getComposites()[0]

# Run the pulse IRTransformation 
optimizer = xacc.getIRTransformation('quantum-control')
optimizer.apply(program, qpu, {
    # Using the Python-contributed pulse optimizer (Krotov)
    'method': 'krotov',
    'max-time': T
})
\end{python}
\caption{Using the pulse \texttt{IRTransformation} service with an external Python plugin (Krotov) as the optimizer. In this example, we optimize for, effectively, an Hadamard gate (expressed as a $XY^{1/2}$ circuit.)}
\label{fig:krotov_snippet}
\end{figure}
\begin{figure}[b!] 
\includegraphics[width=0.5\textwidth]{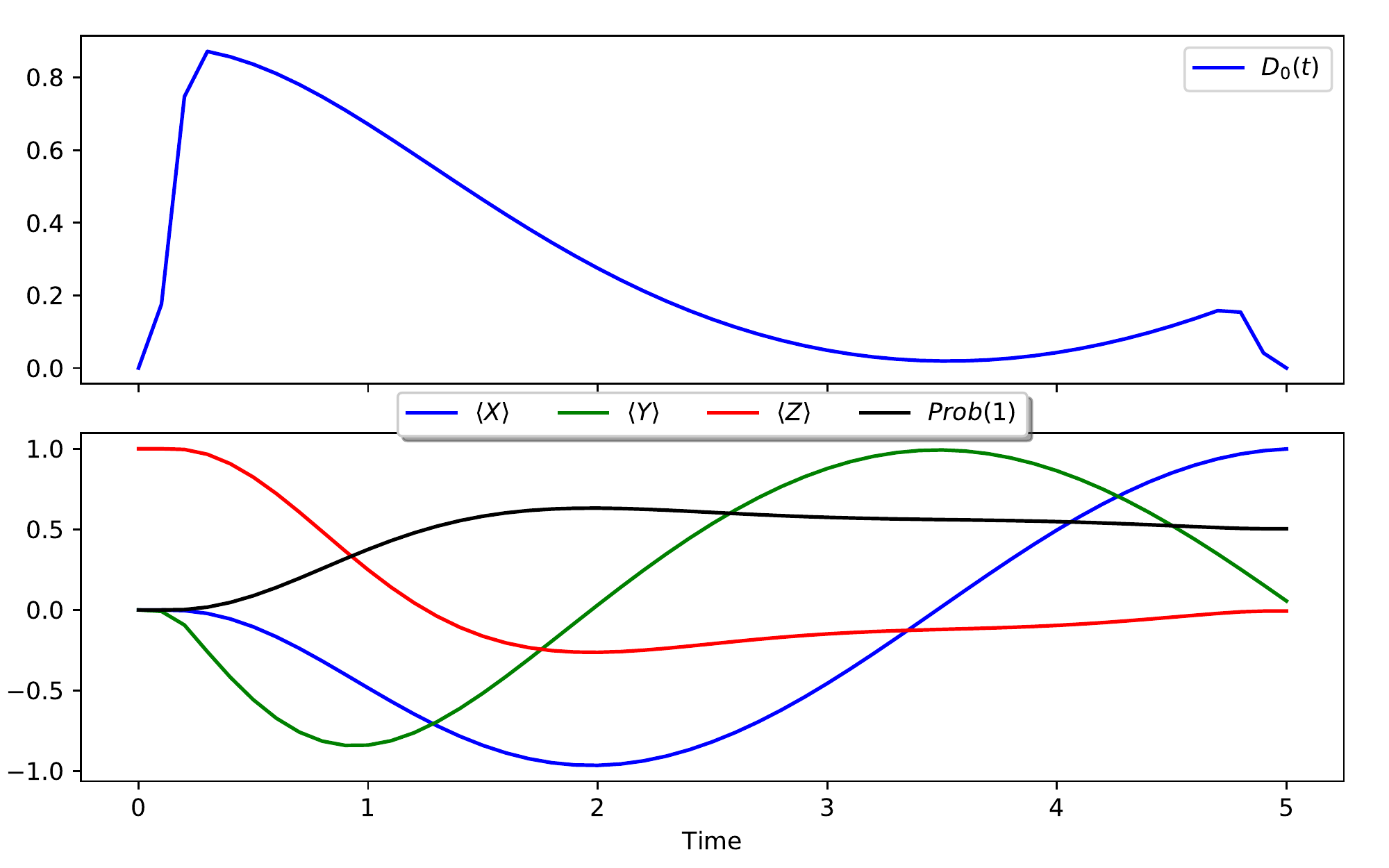}
\caption{Execution results on the QuaC backend for the optimal pulse derived by the \texttt{IRTransformation} service using the Krotov optimizer. The \texttt{IRTransformation} transforms a Hadamard-gate equivalent circuit ($Y^{1/2}-X$) into an analog pulse (\emph{top} panel). The time-domain response on the QuaC backend (expectation values of the number operator and Pauli operators) in the \emph{bottom} panel confirmed a Hadamard response. }
\label{fig:krotov_response}
\end{figure} 
The data generated by the Krotov package while performing the pulse optimization procedure is illustrated in Fig.~\ref{fig:krotov_control_field}. 
Starting with an arbitrary guess pulse, it drives the pulse envelope to an optimal shape via a method-specific update policy to implement the target unitary. It is worth noting that, in this example, we purposely specify the Hadamard gate as a $Y^{1/2}-X$ gate sequence to demonstrate the total unitary computation functionality of the  \texttt{IRTransformation} service. The underlying optimal control module will only receive the computed target unitary matrix as a black box.

The optimal pulse \texttt{IR} after transformation can then be executed on a pulse-capable backend (like lines 27-28 of Fig.~\ref{fig:simple_ir_transform_grape}.) The verification results for the Krotov-optimized pulse on our QuaC simulator backend is shown in Fig.~\ref{fig:krotov_response}. 
The pulse shape, as executed on the backend, is the output from the Krotov optimizer. The expectation values of $X$, $Y$, and $Z$ operators indicate that this pulse indeed performs a Hadamard transformation (the initial state is $|0\rangle$.) 

\begin{figure} [b!]
\includegraphics[width=0.5\textwidth]{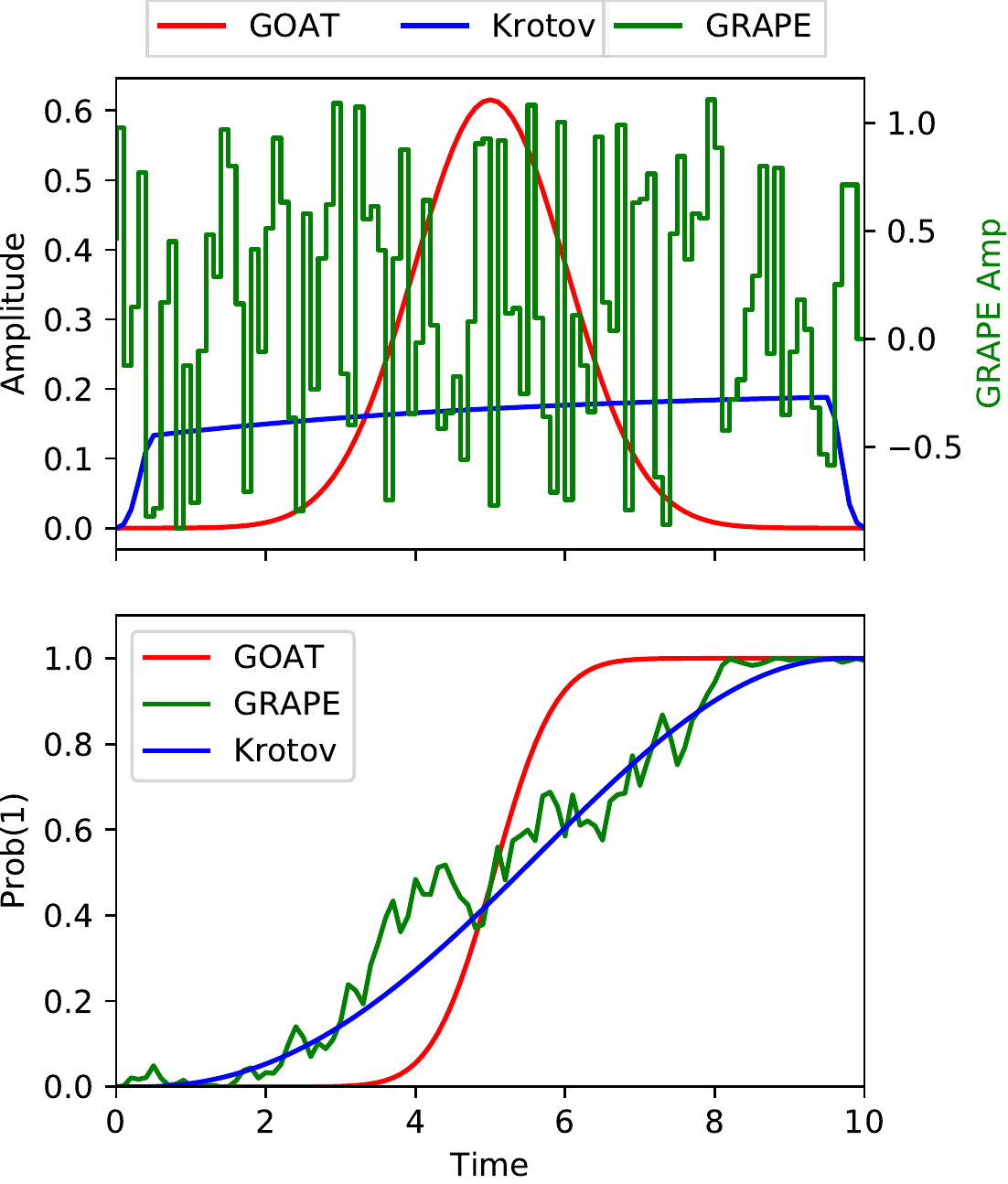}
\caption{Comparison between pulses generated by GOAT, GRAPE, and Krotov optimizers when performing \texttt{IRTransformation} for an $X$ gate: (top) pulse envelopes and (bottom) excited state population (initial state = $|0\rangle$).}
\label{fig:pulse_comparison}
\end{figure} 
\subsection{Comparison Across Techniques}
One key benefit of having multiple pulse optimizers implemented as plugins in a micro-services approach is that users can easily examine different methods in a plug-and-play manner. For instance, we can request an $X$ gate to be transformed into pulses in a similar way to the code snippet in Fig.~\ref{fig:krotov_snippet} (with a different XASM kernel string), in which the method field can be either '\texttt{GOAT}', '\texttt{GRAPE}', or '\texttt{Krotov}'. Correspondingly, the appropriate plugin will be invoked to perform the optimization task with the same set of inputs, i.e. the system Hamiltonian dynamics encapsulated by the \texttt{qpu} instance and the target unitary of the gate-based \texttt{program}.

The generated pulses and the excited state population responses are shown in Fig.~\ref{fig:pulse_comparison}. 
It is worth noting that the generated pulses strongly depend on the initial guess pulses which are Gaussian, square, and random for GOAT, Krotov, and GRAPE, respectively. 

\begin{figure} [b!]
\includegraphics[width=0.5\textwidth]{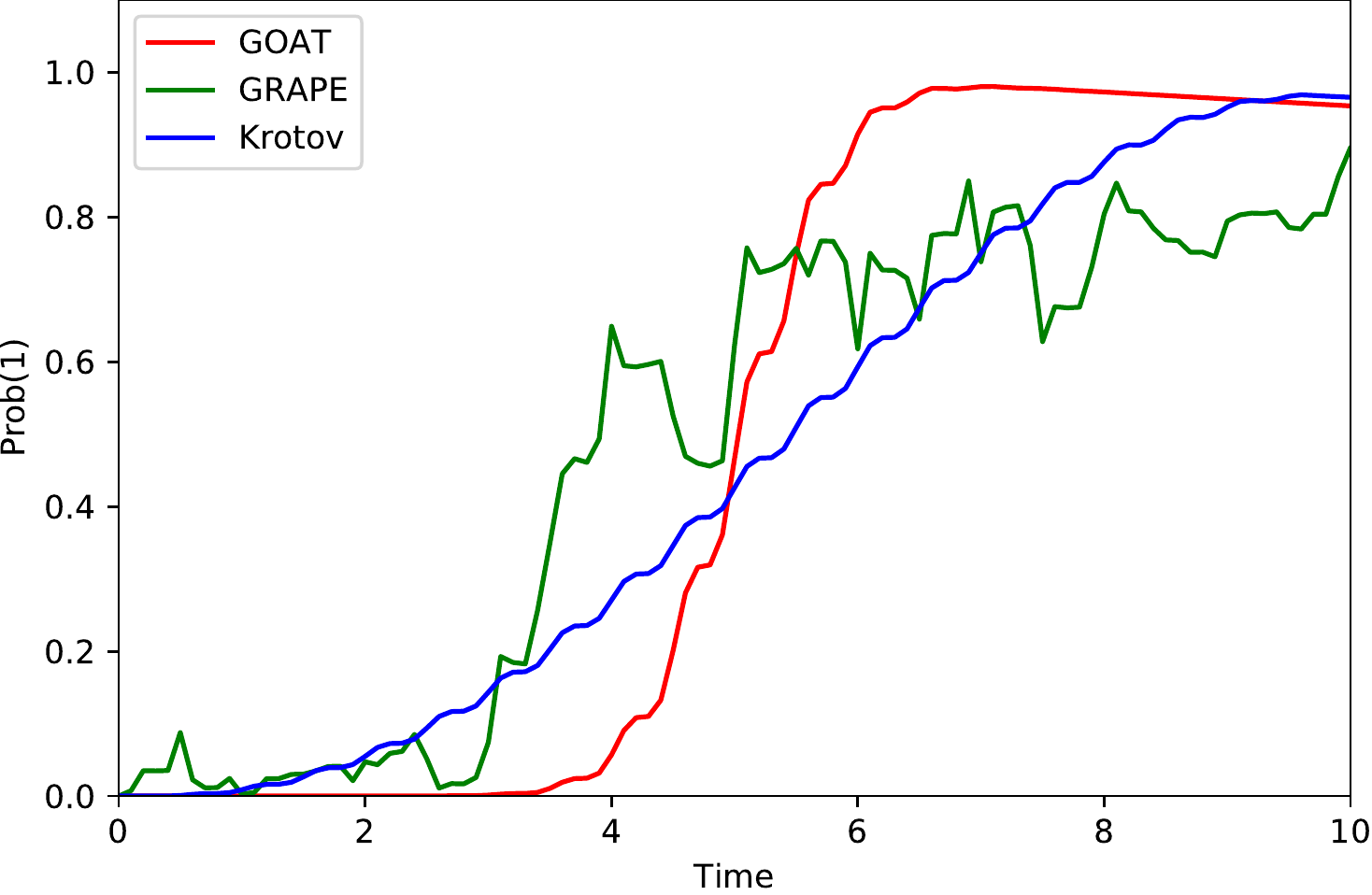}
\caption{Comparison between $X$-gate pulses generated by GOAT, GRAPE, and Krotov optimizers with qubit decay ($T_1 = 10 \times T_{gate}$).}
\label{fig:compare_decay}
\end{figure}

\begin{figure} [b!]
\includegraphics[width=0.5\textwidth]{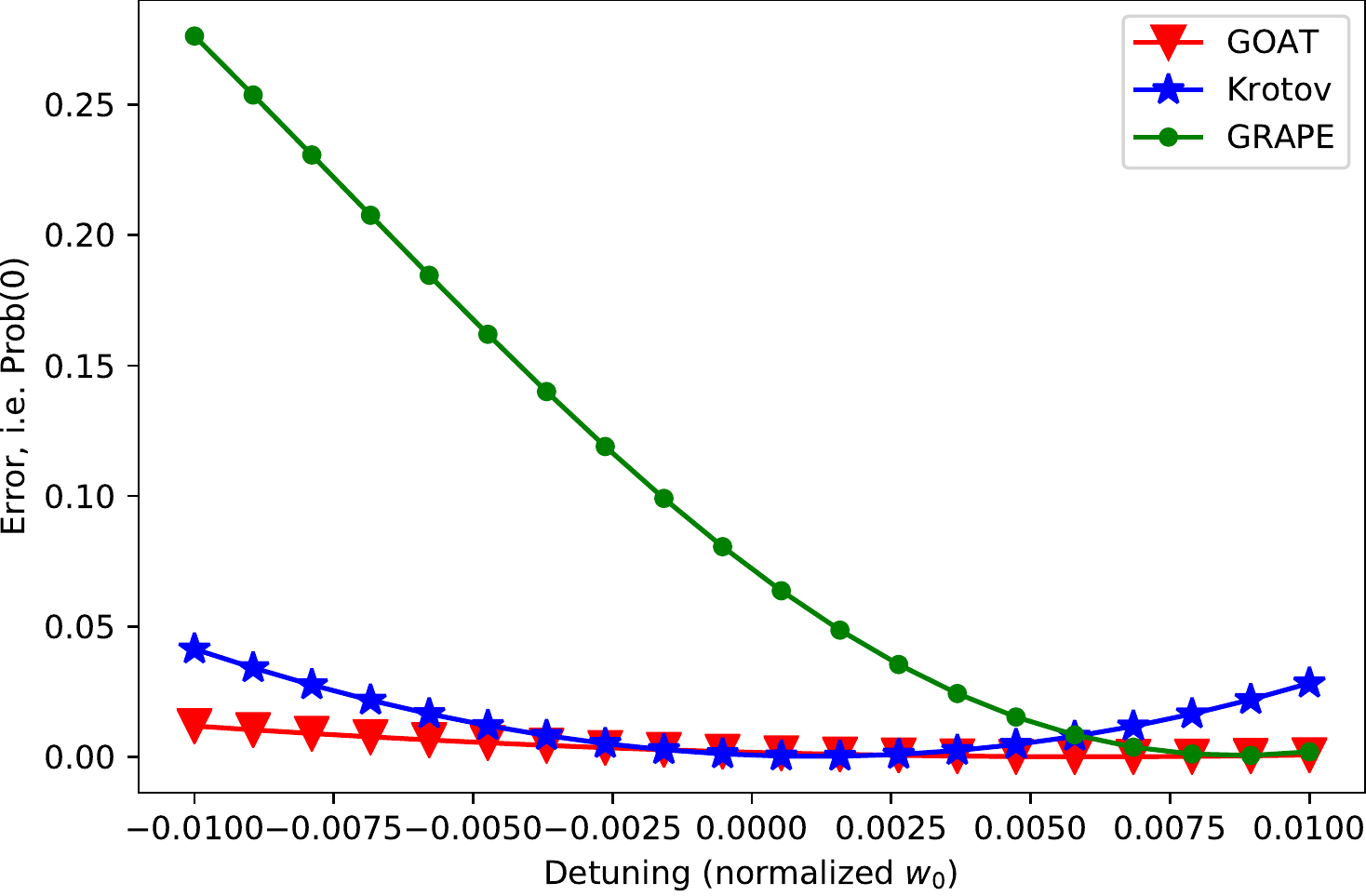}
\caption{Comparison between $X$-gate pulses generated by GOAT, GRAPE, and Krotov optimizers with static detuning ($\omega_{LO} = (1 + \delta)\omega_0$).}
\label{fig:static_detuning}
\end{figure}

Similarly, we can also introduce non-ideal effects such as finite qubit decay \cite{qubitdecay-paper} or local-oscillator detuning \cite{ware2019crossresonance} (from the exact qubit frequency) to the simulator backend and examine the performance of generated pulses under such circumstances. The results for dissipative and detuning experiments are shown in Fig.~\ref{fig:compare_decay} and \ref{fig:static_detuning}, respectively. 
The varying nature of the pulse generated by the randomly-initialized GRAPE method results in loss of fidelity under both the dissipative condition (Fig.~\ref{fig:compare_decay}) and static detuning (Fig.~\ref{fig:static_detuning}).

The results presented here are model- and configuration-dependent but demonstrating the capability and utility of the new quantum control extension to the XACC framework. The modularity and compatibility of XACC services enable users to quickly evaluate different quantum control solutions on a unified API.    

\subsection{Multi-qubit Circuit}
By integrating pulse optimization modules into the IR transformation workflow, as shown in Fig.~\ref{fig:pulse_ir_transformation_flow}, we can transform the whole quantum circuit consisting of multiple gates into a monolithic pulse program implementing the overall unitary. This might be beneficial, e.g., for frequently used sub-circuits because one can derive a unified set of pulses across multiple channels to achieve the overall unitary of the sub-circuit, rather than assembling individual gate pulses. 

In the following example, we use the GRAPE pulse optimizer (Fig.~\ref{fig:python_qft}, line 9) to convert a two-qubit Quantum Fourier Transform (QFT) circuit into a pulse program. 
\begin{figure} [b!]
\begin{python}
# Use XACC QFT circuit generator
qft = xacc.getComposite('qft') 
# Expand QFT circuit for 2 qubits
qft.expand({ 'nq' : 2 })

# Run the pulse IRTransformation 
optimizer = xacc.getIRTransformation('quantum-control')
optimizer.apply(qft, qpu, {
    'method': 'GRAPE',
    'max-time': T,
    'dt': dt
})
\end{python}
\caption{Python code snippet demonstrates the pulse-level IR transformation of a two-qubit Quantum Fourier Transform circuit into a pulse program using the GRAPE method.}
\label{fig:python_qft}
\end{figure} 
It is worth noting that the XACC framework has built-in support for circuit generation of common algorithms. Hence, the QFT circuit (line 1-4 in Fig.~\ref{fig:python_qft}) is automatically expanded to the gate sequence.% in Fig.~\ref{fig:qft_circ}.
% \begin{figure}
% \begin{lstlisting}
% H q1
% CPhase(1.5708) q0, q1
% H q0
% Swap q0,q1
% \end{lstlisting}
% \caption{Quantum circuit that XACC generates for a two-qubit QFT algorithm.}
% \label{fig:qft_circ}
% \end{figure}
\begin{figure}[b!]
\includegraphics[width=0.45\textwidth]{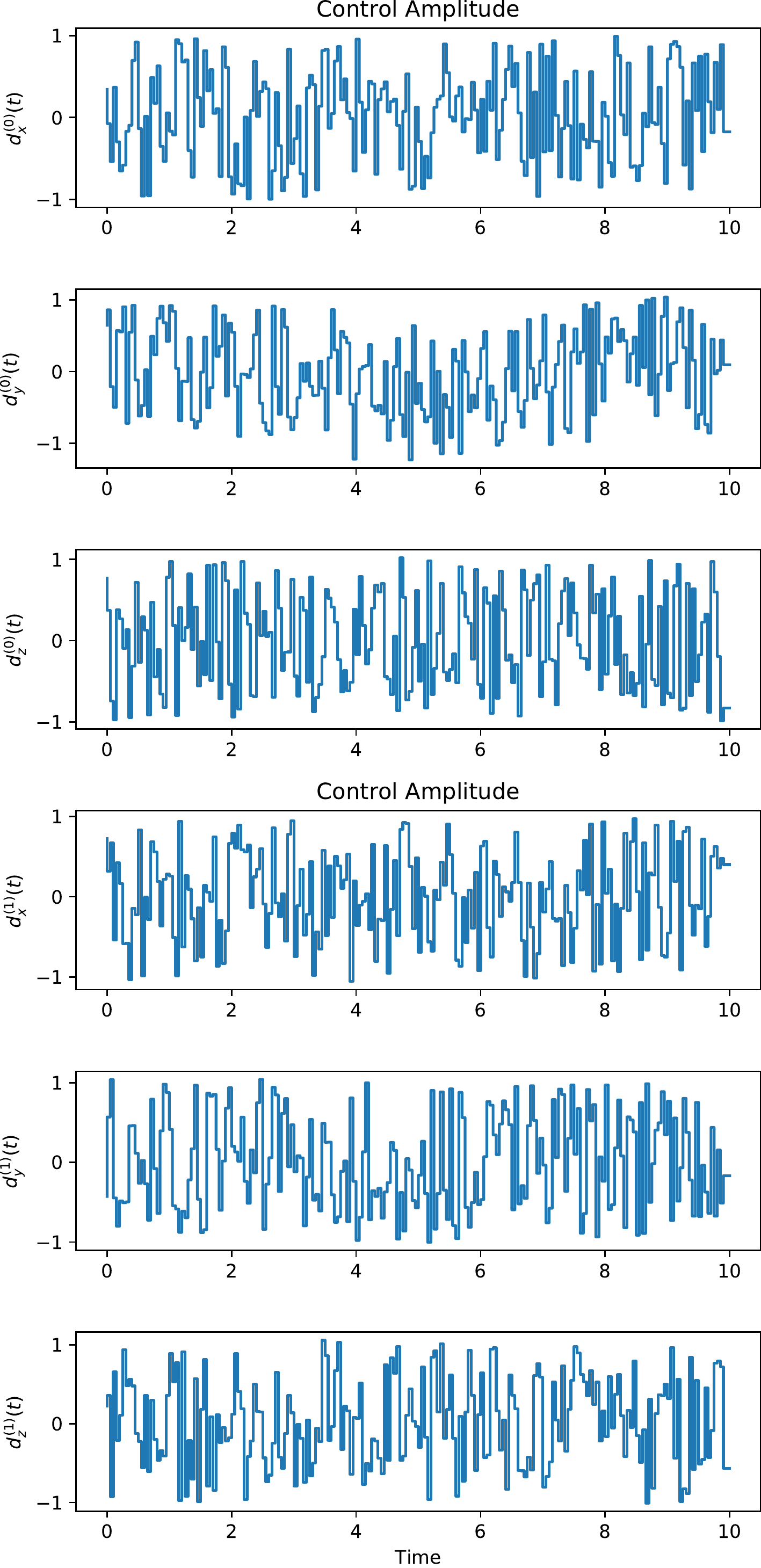}
\caption{Control signals for single-qubit operator terms in Eq.~(\ref{eq:ham_qft_ex}).}
\label{fig:qft_single_channels}
\end{figure} 

In this example, we consider a generic Hamiltonian model in the form  
\begin{equation}
    H = \sum_{k\in\{x,y,z\}}  u_k(t)\sigma_k^{(0)}\sigma_k^{(1)} + \sum_{i= 0}^1\sum_{k\in\{x,y,z\}}d^{(i)}_k(t)\sigma_k^{(i)},
\label{eq:ham_qft_ex}
\end{equation}
where ${u_k(t), d_k^{(i)}(t)}$ are control terms that our optimal control plugin aims to optimize. This simplified model is for illustrative purposes only and does not correspond to any particular physical system.  

The pulse waveforms for ${d_k^{(i)}(t)}$ channels are shown in  Fig.~\ref{fig:qft_single_channels}. Similarly, we also obtained pulses for ${u_k(t)}$ channels which are not shown here for brevity.
This pulse program can then be verified on the QuaC simulator backend, i.e. the \texttt{qpu} instance which was given to the optimizer (Fig.~\ref{fig:python_qft}, line 8). 

\subsection{Demo with physical hardware}
The pulse IR transformation service of XACC also provides a means for users to optimize for particular gates that suit their needs. Hardware vendors usually provide a minimal set of pulses which implement a target universal gate set to which other gates are decomposed. Using the XACC pulse IR transformation service, users can instead opt for a more custom approach as illustrated in Fig.~\ref{fig:rx_pulse_optim}
\begin{figure}[b!] 
\begin{python}
xasmCompiler = xacc.getCompiler('xasm');
ir = xasmCompiler.compile(
'''__qpu__ void Kernel(qbit q, double theta) {
    Rx(q[0], theta);
}''', qpu);

program = ir.getComposites()[0]
for angle in angles:
    evaled = program.eval([angle])
    # Run the pulse IRTransformation 
    optimizer = xacc.getIRTransformation('quantum-control')
    optimizer.apply(evaled, qpu, { <options>})
\end{python}
\caption{Pulse optimization for parametric gates.}
\label{fig:rx_pulse_optim}
\end{figure} 
whereby a parametric rotation gate at a specific angle is directly lowered to a pulse instruction. 

Using IBM Qiskit \cite{Qiskit} 
% \begin{figure} [b!]
% \begin{python}
% for angle in angles:
%     circ = QuantumCircuit(1, 1)
%     circ.rx(angle, 0)
%     circ.measure([0], [0])
%     transpiled_circ = qiskit.transpile(circ, backend)  
%     schedule = qiskit.build_schedule(transpiled_circ, backend)
%     pulse_schedules.append(schedule)
% \end{python}
% \caption{Sample IBM Qiskit code snippet to use the default gate-to-pulse transpiler.}
% \label{fig:qiskit_default_transpile}
% \end{figure}
we can examine the equivalent pulse sequences compiled by the vendor-provided pulse library to implement those parametric gates. These pulses are what will be applied to the underlying hardware when users submit gate-based circuits via XACC, Qiskit, or other front-end tools to the cloud backend.

\begin{figure} 
\includegraphics[width=0.5\textwidth]{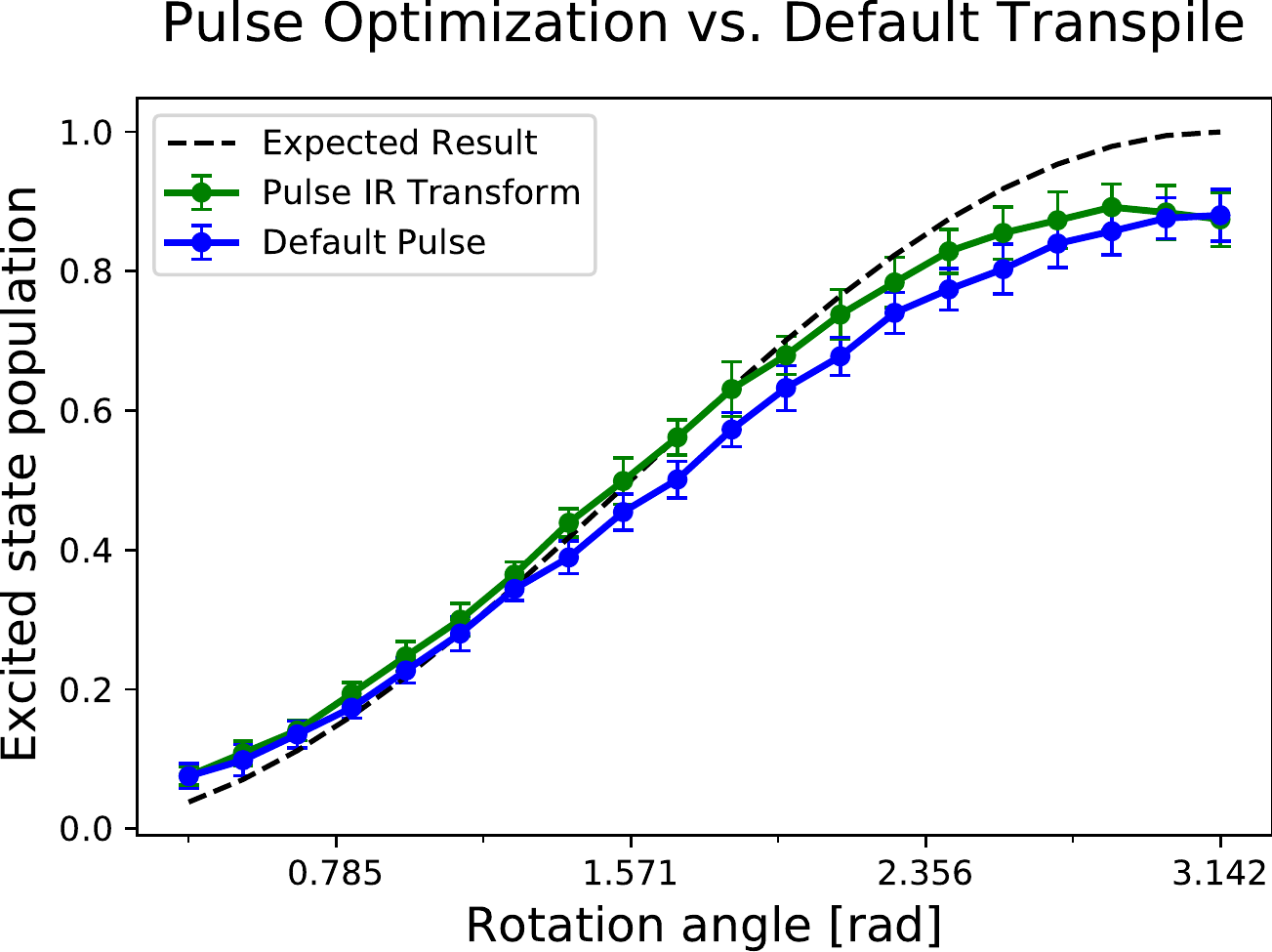}
\caption{Results from running $Rx(\theta)$ gate with different rotation angles on the IBMQ \texttt{armonk} (one qubit) backend. In the runs using XACC pulse optimization (green), rotation gate at a specific angle is transformed into a Gaussian-shaped envelope pulse (GOAT method). In default transpile runs (blue), default Qiskit transpiler is used. For each angle, 1024 shots are used to compute the excited state probability. There are 10 runs for each angle.}
\label{fig:ibm_results}
\end{figure}

Fig.~\ref{fig:ibm_results} 
is a comparison between pulses that are optimized by the XACC GOAT optimizer to implement $R_x(\theta)$ gates at specific $\theta$ values vs. default pulse sequences generated by the Qiskit transpiler for the IBM one-qubit Armonk device. On average, pulses optimized by the XACC optimizer have comparable fidelity to the default ones. When using this IR transformation method for longer gate sequences, we can potentially have a significant gain in terms of execution time and fidelity. 
\section{Discussion and Future Work}
We have expanded upon XACC's pulse-level programming capacity by integrating several common quantum optimal control algorithms into our framework. With an emphasis on ease of implementation, we provide users the ability to take their quantum circuits and compile them to an optimally shaped control pulse. We currently support algorithms such as GRAPE, GOAT, and Krotov, in both C\texttt{++} and Python, with plans to offer more algorithms in the future. Additionally, we hope to exploit the MPI parallelization feature of QuaC in order to use cluster based computing resources such as Oak Ridge's Summit Supercomputer \cite{summit-supercomputer}. This would allow for accurate simulation and subsequent control optimization routines to be run on larger quantum systems. Our long-term focus is both on maintaining our database of calibrated quantum hardware system models, as well as expanding the number of hardware providers supported by our framework for pulse-level control. 

\section*{Acknowledgements}
\label{}
This work has been supported by the US Department of Energy (DOE) Office of Science Advanced Scientific Computing Research (ASCR) Quantum Computing Application Teams (QCAT), Quantum Testbed Pathfinder (QTP), and Accelerated Research in Quantum Computing (ARQC). A.\ S.\ was supported by an appointment to the Oak Ridge Associated Universities (ORAU) Post-Bachelors Program, sponsored by the U.S.\ Department of Energy and administered by the Oak Ridge Institute for Science and Education. ORNL is managed by UT-Battelle, LLC, for the US Department of Energy under contract no. DE-AC05-00OR22725. The US government retains and the publisher, by accepting the article for publication, acknowledges that the US government retains a nonexclusive, paid-up, irrevocable, worldwide license to publish or reproduce the published form of this manuscript, or allow others to do so, for US government purposes. DOE will provide public access to these results of federally sponsored research in accordance with the DOE Public Access Plan.

\bibliographystyle{unsrt}
\bibliography{main}

\end{document}